\documentclass[10pt]{article}
\usepackage{moriond,epsfig}
\usepackage{amssymb,amsmath,mathrsfs}
\usepackage[T1]{fontenc}
\usepackage[applemac]{inputenc}   
\usepackage{mathtools}
\usepackage{url}
\usepackage[pdfstartview = FitH,colorlinks, urlcolor=blue, citecolor=blue, linkcolor=red]{hyperref}
\usepackage{hhline}
\usepackage{tikz}
\usepackage{mathtools}
\usetikzlibrary{decorations}
\usetikzlibrary{decorations.pathreplacing}
\usetikzlibrary{decorations.markings}
\usetikzlibrary{arrows,matrix}
\usepackage{cite,ifpdf}

\def\Journal#1#2#3#4{{#1} {\bf #2}, #3 (#4)}

\def\PRD{{\em Phys. Rev.} D}

\def\be{\begin{equation}}
\def\ee{\end{equation}}
\def\bea{\begin{eqnarray}}
\def\eea{\end{eqnarray}}

\newcommand{\rb}[1]{\raisebox{1.5ex}[0pt]{#1}}
\renewcommand{\thefootnote}{\fnsymbol{footnote}}

\DeclareMathSymbol{\reel}{\mathalpha}{AMSb}{"52}
\DeclareMathSymbol{\zzz}{\mathalpha}{AMSb}{"5A}
\DeclareMathSymbol{\complex}{\mathalpha}{AMSb}{"43}
\DeclareMathSymbol{\tmp}{\mathalpha}{AMSa}{"79}

\newcommand{\pD}{\ensuremath{p_{_{\!D^{**}}}}}%
\newcommand{\pB}{\ensuremath{p_{_{\!B}}}}%
\newcommand{\mD}{\ensuremath{m_{_{\!D^{**}}}}}%
\newcommand{\EB}{\ensuremath{E_{_{\!B}}}}%
\newcommand{\mB}{\ensuremath{m_{_{\!B}}}}%
\newcommand{\poldhh}[3]{\ensuremath{\varepsilon^{#1#2}_{(p_{_{\!#3}}\!)}}}%
\newcommand{\poldbb}[3]{\ensuremath{\varepsilon_{#1#2}^{(p_{_{\!#3}}\!)}}}%

\newcommand{\rDd}{\ensuremath{r_{_{\!D^{*}_2}}}}%
\newcommand{\mDd}{\ensuremath{m_{_{\!D^{*}_2}}}}%
\newcommand{\taub}{\ensuremath{\tau_{_{\!3/2}}}}%

\DeclarePairedDelimiterX\brakket[3]{\langle}{\rangle}{#1\,\delimsize\vert\,#2\,\delimsize\vert\,#3}%
\DeclarePairedDelimiterX\braket[2]{\langle}{\rangle}{#1\,\delimsize\vert\,#2}%
\DeclarePairedDelimiterX\bra[1]{\langle}{\rvert}{#1}%
\DeclarePairedDelimiterX\ket[1]{\lvert}{\rangle}{#1}%

\newcommand{\etatp}[2]{\ensuremath{\prescript{#1}{}{P}_#2}}%
\begin{document}
\tikzset{->-/.style n args={2}{decoration={markings, mark=at position #1 with {\arrow{#2}}},postaction={decorate}}}
\vspace*{4cm}
\title{\Large\sc{Lattice computation of ${B \to D^*,\;D^{**}\ell \nu}$ form factors \\ at finite heavy masses}}
\author{ \normalsize {\bf{{ Mariam~Atoui}}} \footnote{email adress: \href{mailto:matoui@in2p3.fr}{\sf matoui@in2p3.fr}} \footnote{CNRS Liban scholarship}}\address{
\vskip 0.5cm
Laboratoire de Physique Corpusculaire, Université Blaise Pascal, CNRS/IN2P3 \\ 63177 Aubière Cedex, France\\[2mm]
}
\maketitle
\vskip 1cm
\abstracts{\sf \normalsize{We propose a strategy to compute form factors entering the semileptonic decay channel of $B$ mesons into orbitally excited (P wave) $D^{**}$ charmed mesons on the lattice using, for the first time, realistic charm quarks having a finite mass. We present preliminary results about the extracted transition amplitudes and form factors at different recoils and at three different $b$ quark masses.}}

\newpage
\section{Introduction}
\setcounter{footnote}{0}
\def\thefootnote{\arabic{footnote}}
The phenomenology of $b$ hadrons is very rich, because owing to the large mass of the bottom quark there are many decay channels. Available semileptonic decays into any hadronic part containing a charm quark are among the decay modes of $B$ mesons which play a critical role in the analysis of unitarity triangle and are an essential ingredient to any precise information about the CKM matrix element $|V_{cb}|$.\\
\noindent However, there are many puzzling features associated with the semileptonic $b\to c$ data which have existed during the last ten years as, for example, the so called ``1/2 versus 3/2 puzzle" \cite{Bigi:2007qp,LeYaouanc:2001nk}. It corresponds to the persistent conflict between theoretical predictions and experimental measurements of semileptonic branching ratios of $\bar{B}\to D^{**} \ell \nu$:
\begin{align*}
\Gamma (\bar{B}_d \to D^{**}_{{1/2}_{\text {broad}}} \ell \nu) &\ll \Gamma(\bar{B}_d \to D^{**}_{{3/2}_{\text {narrow}}} \ell \nu) \qquad [\text{Theory}] \\
\Gamma(\bar{B}_d \to D^{**}_{{1/2}_{\text {broad}}} \ell \nu) &\gg \Gamma(\bar{B}_d \to D^{**}_{{3/2}_{\text {narrow}}} \ell \nu)  \qquad  [\text{Experimental data}]
\end{align*}
where $D^{**}$ are the first orbital excitations of $D$ mesons having a positive parity.
In the quark model picture, if we consider $P$ wave states we see that four orbital excitations of heavy mesons should exist. In the case where the heavy quark is the charm quark $c$, we obtain the $D^{**}$ represented in table~\ref{D**}:
\begin{table}[h]
\begin{center}
\begin{tabular}{||c||c||c||}
\hhline{|t:=:t:=:t:=:t|}
doublet&  $J^P$ values & experimental notation\\
\hhline{|:=::=::=:|}
& $0^+$ & $D_0^*$\\
\hhline{||~||-||-||}
 \rb{$j^P=1/2^+$} &$1^+$ & $D_1^*$\\
\hhline{|:=::=::=:|}
& $1^+$ & $D_1$\\
\hhline{||~||-||-||}
 \rb{$j^P=3/2^+$}&$2^+$ & $D_2^*$\\
\hhline{|b:=:b:=:b:=:b|}
\end{tabular}
\caption{\sf \label{D**} Classification of $D^{**}$ states. The  total angular momentum $J=j+s_c$  is the sum of $j$ (the angular momentum of the light component of a heavy meson) and $s_c=1/2$ (spin of the heavy quark). The first column contains the classification in the infinite mass limit where due to the heavy quark symmetries, $j$ is a conserved quantity and thus a good quantum number to classify the states.}
\end{center}
\end{table}
\noindent So, the comparison between experimental and theoretical results presents some hardships and this challenges our understanding of QCD and is of high relevance for an accurate estimate of $V_{cb}$.\\
However, in many theoretical approaches (HQET, heavy quark expansion, quark model, Lattice QCD with quenched approximation, etc...), branching ratios corresponding to the $\bar{B}\to D^{**} \ell \nu$ decay were calculated using the infinite mass limit. That is the reason why, in order to address the aforementioned questions, we propose to determine the form factors and then the branching ratios using ``real" charmed quarks having a finite mass.\\
\noindent In the following, I will explain the extraction of form factors corresponding to the decay of $B\to D^{**}$ in the framework of Lattice QCD and I will discuss some preliminary results we have obtained. This work is done in collaboration with B.~Blossier and O.~Pène ({\em{LPT Orsay France}}),  V.~Morénas ({\em{LPC Clermont Ferrand France}}) and K.~Petrov ({\em{LAL Orsay France}}).
\section{Form factors\label{ff}}
We choose to consider in our work the decay of $B$ into the scalar {\em $\etatp{3}{0}$} and the tensor {\em $\etatp{3}{2}$} states of $D^{**}$. The semileptonic decay of a pseudoscalar meson into a scalar meson is mediated by the axial part of the weak $({V-A})_\mu$ current. The operator $V_\mu$ denotes the vector current $\bar q\gamma_\mu b$ and the operator $A_\mu$ represents the axial current $\bar q\gamma_\mu\gamma_5 b$.\\
The matrix element for the $\etatp{3}{0}$ state can be parametrized in terms of two form factors ($\tilde u_+, \tilde u_-$):
\begin{equation*}
\brakket[\big]{\etatp{3}{0}(\pD)}{A_\mu}{B(\pB)}=
\boxed{\tilde u_+}\,(\pB+\pD)_{\mu} + 
\boxed{\tilde u_-}\,(\pB-\pD)_{\mu}
\label{eq:3P0}
\end{equation*}
and for  $\etatp{3}{2}$ state: (Here, $\lambda$ is the polarization tensor corresponding to $J=2$ states, $\lambda=\{0,\pm 1,\pm 2\}$)
\[
\begin{aligned}
\brakket[\big]{\etatp{3}{2}(\pD,\,\lambda)}{V_\mu}{B(\pB)}\ &=\ i\,
\boxed{\tilde h}\,\epsilon_{\mu\rho\sigma\tau}\,
\poldhh{\rho}{\alpha\ast}{D^{\ast\ast}}\,{\pB}_{\!\alpha}\,
(\pB+\pD)^{\sigma}\,
(\pB-\pD)^{\tau} \\
\brakket[\big]{\etatp{3}{2}(\pD,\,\lambda)}{A_\mu}{B(\pB)} &= \boxed{\tilde k}\,\poldbb{\mu}{\rho}{D^{\ast\ast}}\pB^{\rho} \\
&+
\left({\poldbb{\alpha}{\beta}{D^{\ast\ast}}\pB^{\alpha}\pB^{\beta}}\right)
\left[{\boxed{\tilde b_{+}}\,(\pB+\pD)_{\mu} +
\boxed{\tilde b_{-}}\,(\pB-\pD)_{\mu}
}\right]
\end{aligned}
\]

\section{Going to the Lattice}
The determination of many observables such as form factors, decay constants as well as numerous matrix elements, which play an important role in Flavor Physics, requires the use of non perturbative methods because when we look at the low properties of QCD, we can no longer use perturbative techniques   \footnote{At short distances or at high energies the quarks interact weakly, so that it is possible to study the theory of strong interactions (QCD) with perturbative techniques since the coupling constant $\alpha_{_{\text{QCD}}} \ll 1$}. Lattice QCD \footnote{For further reading, I refer the reader to \cite{book1,book2} } is considered as the only way to systematically and rigorously solve non perturbatively the quantum theory of strong interactions starting from first principles.\\
Briefly, it is a means of regularizing Field Theory where the continuum and infinite space-time is replaced with a discretized grid of points in a finite volume of extent $L$ in space and $T$ in time, separated by a distance $a$ (i.e. the lattice spacing). Quark fields are present on the sites of the lattice and gauge fields are the links between those sites.

\subsection{LQCD action}
\noindent The {\bf{\em gauge action}} used in our simulation is tree-level Symanzik improved \cite{gaugeact} with $\beta=3.9$ corresponding to a lattice spacing $a=0.0855$ fm and
where $b_1=-1/12$, $b_0=1-8b_1$.\\
The {\bf{\em fermionic action}} is the Wilson Twisted-mass Lattice QCD (tmLQCD) action with two flavors of mass-degenerate quarks, tuned at {\em maximal twist} in the way described in full details in Ref. \cite{tm}:
\subsection{Simulation setup}
We have computed green functions using two ensembles of gauge configurations produced by the European Twisted Mass Collaboration (ETMC). Simulation parameters are presented in Table~\ref{parameters}.
\begin{table}[here]
\begin{center}
\begin{tabular}{||c||c||c||c||c||c||}
\hhline{|t:=:t:=:t:=:t:=:t:=:t:=:t|}
$\beta$ &$L^3\times T$ & $\mu_{\text{sea}}=\mu_l$ & $\mu_c$ & $\mu_h$ & nb. of gauge configurations\\
\hhline{|:=::=::=::=::=::=:|}
3.90 & $24^3 \times$ 48 & 0.0085 &0.215 & 0.3498, 0.4839, 0.6694 & 240\\
4.05 & $32^3 \times$ 64 & 0.0060 &0.1849 & 0.3008, 0.4162, 0.57757 & 200\\
\hhline{|b:=:b:=:b:=:b:=:b=:b:=:b|}
\end{tabular}
\caption{\sf \label{parameters} Parameters we have considered in our simulation. $\mu_l, \mu_c, \mu_h$ are respectively the bare twisted light, charm and heavy quark masses}
\end{center}
\end{table}
\section{Determination of form factors}
The first step in the determination of form factors, presented in section~\ref{ff}, is the computation of matrix elements contributing to these form factors, and in order to access matrix elements on the lattice one computes the following three-point correlation functions
\begin{figure}[h!]
\[
\begin{minipage}{110mm}
\begin{center}
\begin{tikzpicture}[>\relax=latex,scale=1]
\draw [dashed] (0,-1.25) -- (0,1.25);
\draw [dashed] (3.5,-1.25) -- (3.5,1.25);
\draw [dashed] (1.75,-1.25) -- (1.75,1.25);
\draw (0,-1.25) node [below] {\scalebox{0.8}{$t_i$}};
\draw (0,1.25) node [above] {\scalebox{0.6}{(source)}};
\draw (3.5,-1.25) node [below] {\scalebox{0.8}{$t_f$}};
\draw (3.5,1.25) node [above] {\scalebox{0.6}{(sink)}};
\draw (1.75,-1.25) node [below] {\scalebox{0.8}{$t$}};
\draw [thick,->-={0.5}{>}] (3.5,0) .. controls (3,0.8) and (2.0,0.8) .. (1.75,0.8) node [above, pos=0.4] {\scalebox{0.9}{$\bar c$}};
\draw [thick,->-={0.5}{>}] (1.75,0.8) .. controls (1.5,0.8) and (0.5,0.8) .. (0,0) node [above, pos=0.6] {\scalebox{0.9}{$\bar b$}};
\draw [thick,->-={0.75}{>}] (0,0) .. controls (0.75,-1.) and (2.75,-1) .. (3.5,0) node [below, pos=0.75] {\scalebox{0.9}{$q$}};
\filldraw[draw=black,fill=black] (1.65,0.7) rectangle (1.85,0.9);
\draw [<-\relax] (1.60,0.9) .. controls (1.2,0.95) and (1.4,1.3) .. (1.75,1.4) node {\scalebox{0.8}{$\quad J_\mu$}};
\draw[fill=black] (0,0) circle (1mm);
\draw[fill=black] (3.5,0) circle (1mm);
\draw (-0.05,0) node [left] {\scalebox{0.8}{${\vec x}_i$}};
\draw (-0.05,-0.1) node [below left] {\scalebox{0.8}{$({\vec p}_i)$}};
\draw (3.55,0) node [right] {\scalebox{0.9}{${\vec x}_f$}};
\draw (3.55,-0.1) node [below right] {\scalebox{0.8}{$({\vec p}_f)$}};
\draw (1.75,0.8) node [below right] {\scalebox{0.9}{${\vec x}$}};
\draw (1.75,0.5) node [below right] {\scalebox{0.8}{$({\vec p})$}};
\draw (0,0) node [right] {\scalebox{0.9}{$ {\bar B}$}};
\draw (3.5,0) node [left] {\scalebox{0.9}{${D^{**}}$}};
\end{tikzpicture}
\caption{\sf Sketch of the valence quark flow in the form factor of $B\to D^{**} \ell \nu$}
\end{center}
\end{minipage}
\]
\end{figure}
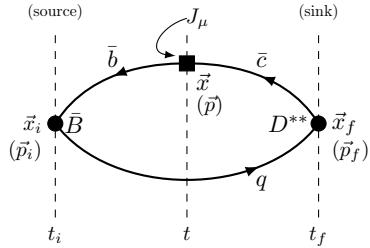
\begin{eqnarray*}
{\mathscr{C}}^{(3)}(t,t_i,t_f,\vec{p}_i,\vec{p}_f) =  \sum_{\text{positions}} \langle { \mathscr{O}}^\dagger_{D^{**} } (t_f,\vec{x}_f) \; J_\mu (t,\vec{x})\; {\mathscr{O}}_B(t_i,\vec{x}_i) \rangle
\cdot e^{i(\vec{x}-\vec{x}_f)\cdot \vec{p}_f} \cdot e^{-i(\vec{x}-\vec{x}_i)\cdot \vec{p}_i}
\end{eqnarray*}
where $\mathscr{O}^\dagger_{D^{**}}$, ${\mathscr{O}}_B$ are respectively the creation and annihilation operators of $D^{**}$ and $B$ mesons, $J_\mu$ is the vector or axial current.\\
From the asymptotic behavior of the three-point correlation function, it is clear that the removal of the exponential factors can be achieved by considering the ratio
\begin{eqnarray}
\mathscr{R}(t) &= \dfrac{\mathscr{C}^{(3)} (t,t_i,t_f,\vec{p}_i,\vec{p}_f)}{\mathscr{C}^{(2)}_{(B)}(t-t_i,\vec{p}_f)\cdot \mathscr{C}^{(2)}_{(D)}(t_f-t,\vec{p}_i)}
\cdot \sqrt{\mathscr{Z}_B}  \cdot \sqrt{\mathscr{Z}_D} 
\label{ratio}
\end{eqnarray}
where $ \mathscr {Z}_M  = \; |\langle 0 | \mathscr{O}_M| M \rangle|^2$ is obtained from the fit with asymptotic behavior of the two-point correlation functions.
\begin{equation*}
\mathscr{C}^{(2)} (t) \xrightarrow[{t\to \infty}]{} \dfrac{\mathscr{Z}_M}{2E_M}\;e^{-E_Mt}
\end{equation*}
When the operators in the ratio \eqref{ratio} are sufficiently separated in time, one observes the stable signal (plateau), which is the desired hadronic matrix element:
\begin{equation*}
\mathscr{R}(t) \xrightarrow[{t-t_i \to \infty}]{t_f-t \to \infty }  \,  \langle D^{**}_{(\vec{p}_f)} | (A_\mu, \, V_\mu) | B_{(\vec{p}_i)} \rangle 
\end{equation*}
\subsection{$\tilde{k}$ form factor}
In order to obtain simpler expressions representing the interpolating fields \footnote{interpolating fields are the mesonic creation and annihilation operators which defines the quantum number of a state. To find the interpolating fields of $2^+$ states, we followed a strategy based on group theory where non local operators are implemented~\cite{group}.} of the P wave excited states, we choose to work in the rest frame of $D^{**}$ mesons where $\pD=(\mD,\vec{0})$ and we also choose a symmetric momentum for the $B$ meson $\pB=(\EB,p,p,p)$.  Using different values of the recoil parameter $w=\tfrac{E_B}{m_B} \; \in \{1,\;1.025,\;1.05,\;1.1,\;1.15,\;1.2,\;1.3 \}$, we perform the extraction of $\tilde{k}$ which is the form factor contributing the most to the decay of $B$ into $D^{**}$.\\
Combining the three point correlation functions contributing to $\tilde{k}$
 \footnote{For example, if we consider the $\lambda=0$ polarization of the $\etatp{3}{2}$ state, $\tilde{k}$ could be written as: 
\[ -\tfrac{\sqrt{6}}{p}  \brakket[\big]{\etatp{3}{2}(0)}{A_1}{B(\pB)}\]
}
, we study the difference with respect to the infinite mass limit ones:
\[{\mathscr{C}^{(3)}_{\text{\tiny{infinite mass limit}}} = p\; {{\tilde{k}_{_\infty}}}\, \cdot \dfrac{\mathscr{C}^{(2)}_B\, \mathscr{C}^{(2)}_D }{\sqrt{\mathscr{Z}_B \mathscr{Z}_D}}}\]
where $\tilde{k}_{_\infty}$ is the form factor at the infinite mass limit. Using what was found in \cite{Morenas:1997nk}:
\[
{{\tilde{k}_{_\infty} }}= \sqrt{3}\,\sqrt{\smash[b]{\rDd}}\,(1+w)\,\taub(w)
 \quad\quad \mDd = \rDd\,\mB
 \]
\[
\taub(w) = \taub(1)\,\left(\tfrac{2}{1+w}\right)^{2\,\sigma^2_{_{\!3/2}}} \quad\quad\quad \taub(1)\simeq 0.539 \quad\text{and}\quad \sigma^2_{_{\!3/2}} \simeq 1.50
\]
we find that for  $w=1.3$ (using for the moment $\beta=3.90$):
\[
\dfrac{\mathscr{C}^{(3)}_{\text{\tiny{finite mass}}}}{\mathscr{C}^{(3)}_{\text{\tiny{infinite mass limit}}}} =
\begin{cases} 
a\mu=0.34 &\leadsto \; 1.38 \pm 0.49\\
a\mu= 0.45 &\leadsto \; 1.97 \pm 0.65\\
{{a\mu=0.67}} &\leadsto \; {{2.88 \pm 3.82}} 
\end{cases}
\]
The correlators corresponding to the higher $b$ mass are characterized by large fluctuations in the effective mass plateaus and thus by large statistical uncertainties. Taking these results with a pinch of salt, and excluding data of the highest heavy quark mass from the next analysis, we estimate roughly the branching fractions $\mathscr{B}(B\to D_{_{(2^+)}}) \cdot \mathscr{B}(D_{_{(2^+)}} \to D^* \pi)$. 
\[
\mathscr{B}(B\to D_{_{(2^+)}}) \cdot \mathscr{B}(D_{_{(2^+)}} \to D^* \pi) = 
\begin{cases} 
m_{B _{\text {GeV}}} = 2.4  & \leadsto (3.6 \pm 2.6) \times 10^{-3} \\
m_{B _{\text {GeV}}} = 2.9  & \leadsto (7.4  \pm 4.9) \times 10^{-3}
\end{cases}
\]
Although we are still far from the physical $B$ mass, we have an indication that there are many observables in $B$ physics that can be determined using LQCD. When results are more refined we aim to do a more detailed comparison with the experimental data. Results will be published as soon as ready.

\subsection{Scalar transition amplitude}
The determination of matrix elements corresponding to the decay of $B$ into $\etatp{3}{0}(\pD)$ state, at different recoils $w$, leads to the weak form factors $\tilde u_+$ and $\tilde u_-$.
For the moment, we focus our attention on the extraction of $\brakket[\big]{\etatp{3}{0}(\pD)}{A_0}{B(\pB)}$ at zero recoil. It seems that this hadronic matrix element is not equal to zero contrary to what was already found in the infinite mass limit.
 \begin{figure}[here]
 \begin{minipage}[b]{.55\linewidth}
 \begin{center}
\includegraphics[width=16pc]{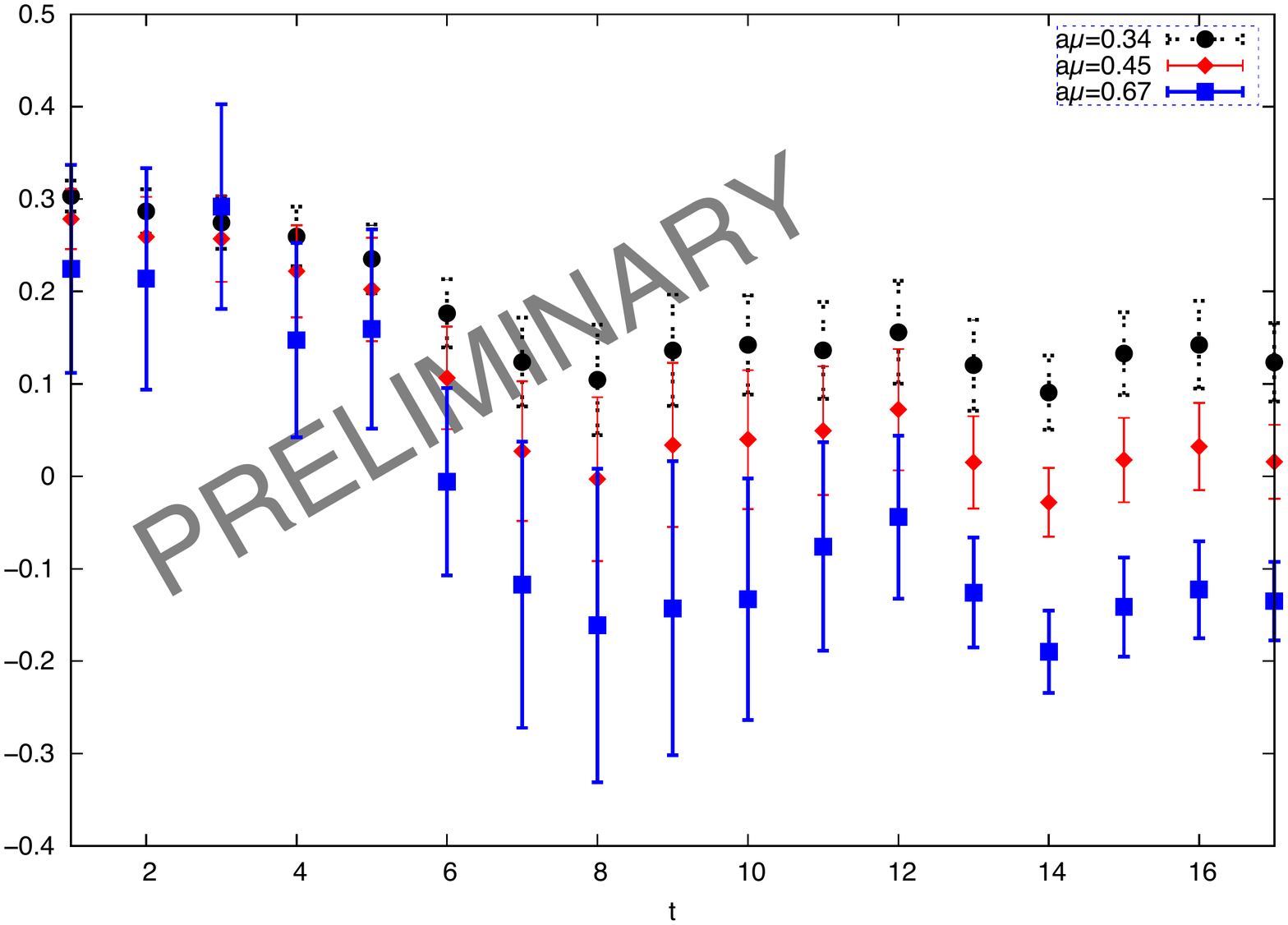}
 \vspace{-1pc}
 \caption{\label{BtoDscab3.9}$\brakket[\big]{\etatp{3}{0}(\mD)}{A_0}{B(\mB)}$ \sf{{at $\beta=3.9$ }}}
 \end{center}
 \end{minipage}\hspace{-2pc}\hfill \begin{minipage}[b]{.55\linewidth}
 \begin{center}
 \includegraphics[width=16pc]{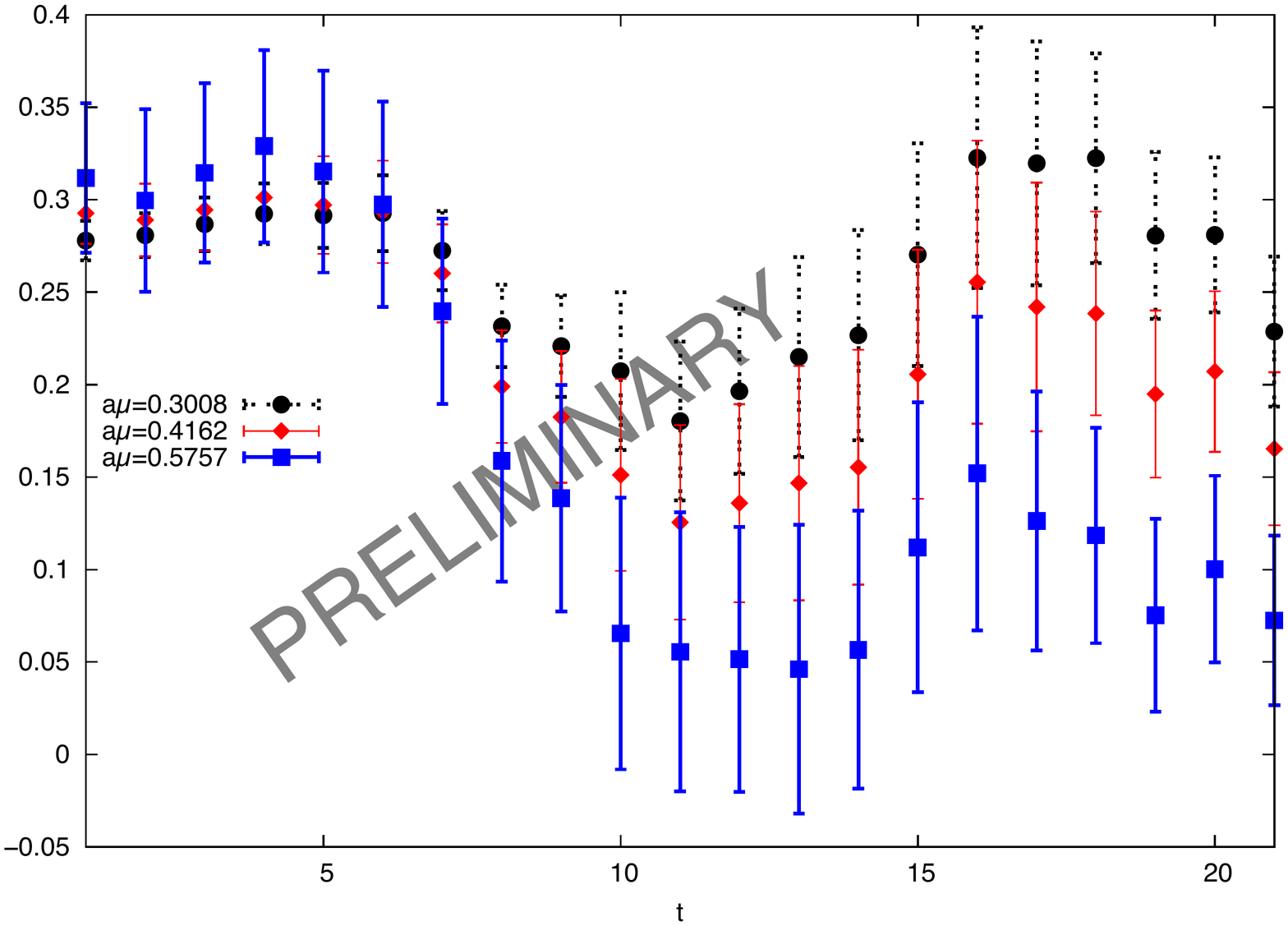}
 \vspace{-1pc}
\caption{\label{BtoDscab4.05}$\brakket[\big]{\etatp{3}{0}(\mD)}{A_0}{B(\mB)}$  \sf{{at $\beta=4.05$}}}
 \end{center}
 \end{minipage}
\end{figure}
\noindent Note that it is an important result that the matrix element $\brakket[\big]{\etatp{3}{0}(\mD)}{A_0}{B(\mB)}$ is non vanishing. It shows us how working with a realistic charm quark could lead to much more accurate results than working with the infinite mass limit approximation.

\subsection{$\mathscr{F}_0(1)$ at zero recoil}
\begin{figure}[here]
 \begin{minipage}[b]{.99\linewidth}
 \begin{center}
\includegraphics[width=20pc]{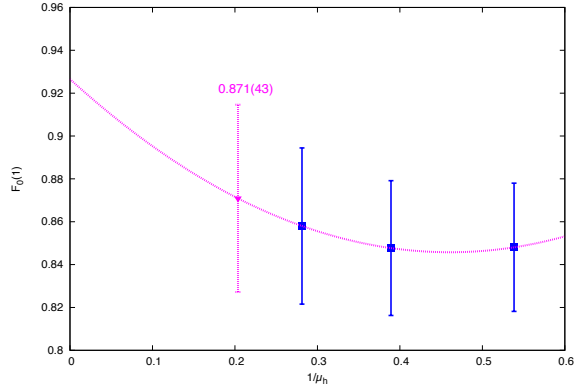}
\vspace{-1cm}
\caption{ \label{FDstar} Extrapolation of $\mathscr{F}_0(1)$ at finite physical $b$ quark masses}
 \end{center}
 \end{minipage}
 \end{figure}

As a byproduct of our analysis, we can also study the $B\to D^* \ell \nu$ decay at zero recoil. The corresponding form factor reads\cite{formfact,formfact2}:
\[
\mathscr{F}_0(1)=Z_A\; \dfrac{\brakket[\big]{D^*(p_{D^*})}{A_i}{B(\pB)}}{2\sqrt{m_B\;m_{D^*}}}
\]
where $Z_A$ is the axial renormalisation constant taken from \cite{renorm}. The values obtained for $\mathscr{F}_0(1)$ are presented in tables~\ref{ffdstar1} and~\ref{ffdstar2}. They do not show any contradiction with what was already found using Lattice QCD with a different fermionic action and different approximations, but the extrapolated value (Fig.~\ref{FDstar}) seems to be higher than what was already found using QCD sum rules \cite{gambino}.
 \begin{table}[here]
 \begin{minipage}[b]{.55\linewidth}
 \begin{center}
 \begin{tabular}{||c||c||}
 \hhline{|t:=:t:=:t|}
 \raisebox{-0.3ex}{  $a\mu_b$} & \raisebox{-0.3ex}{ {$\mathscr{F}_0(1)$} } \\[2mm]
  \hhline{|:=::=:|}
   \raisebox{-0.3ex}{   0.35} &  \raisebox{-0.3ex}{0.882(44)} \\[2mm]
\hhline{|:=::=:|}
   \raisebox{-0.3ex}{  0.48}& \raisebox{-0.3ex}{0.874(48)} \\[2mm]
 \hhline{|:=::=:|}
  \raisebox{-0.3ex}{   0.67} &  \raisebox{-0.3ex}{0.888(58)} \\[2mm]
 \hhline{|b:=:b:=:b|}
 \end{tabular}
\caption{ \label{ffdstar1} ${\mathscr{F}_0(1)}$  \sf{{at $\beta=3.9$ for different values of $a\mu_b$}}}
 \end{center}
 \end{minipage}\hspace{-2pc}
\begin{minipage}[b]{.55\linewidth}
 \begin{center}
 \begin{tabular}{||c||c||}
 \hhline{|t:=:t:=:t|}
 \raisebox{-0.3ex}{  $a\mu_b$} & \raisebox{-0.3ex}{ {$\mathscr{F}_0(1)$} } \\[2mm]
  \hhline{|:=::=:|}
   \raisebox{-0.3ex}{   0.301} &  \raisebox{-0.3ex}{ 0.814(48)} \\[2mm]
\hhline{|:=::=:|}
   \raisebox{-0.3ex}{  0.416}& \raisebox{-0.3ex}{0.821(50)} \\[2mm]
 \hhline{|:=::=:|}
  \raisebox{-0.3ex}{   0.5757} &  \raisebox{-0.3ex}{0.827(53)} \\[2mm]
 \hhline{|b:=:b:=:b|}
 \end{tabular}
\caption{ \label{ffdstar2} ${\mathscr{F}_0(1)}$  \sf{{at $\beta=4.05$ for different values of $a\mu_b$}}}
 \end{center}
 \end{minipage}
\end{table}
 
\section{Discussion}
During the last ten years, techniques and algorithms in LQCD have been developed and computing power has increased: this is really encouraging for the Lattice community where some important parameters in $B$ physics are and will be calculated reaching a precision on par with the experimental one.\\
However, working with excited states on the lattice is not a trivial issue. On one side, analytical expressions of $B\to D^{**}$ transition amplitudes, form factors and decay rates can be calculated. On the other side, isolation of such excited states is very delicate and there is an increase in noise when going to high momentum and discretization errors become larger when the $b$ quark mass increases. We hope that with higher statistics, results will be more clear in order to extract the sought form factors and to discuss their implications.
 
\section*{Acknowledgments}
Mariam Atoui wants to thank the Lebanese National Center for Scientific Research CNRS Liban for the financial support  and acknowledges the help from the Lebanese University during the first year of her PhD thesis.


\end{document}